\documentclass{aa}
\usepackage{graphicx}
\begin{document}
   \title{Cluster AgeS Experiment. Catalog of Variable Stars in the
          Globular Cluster $\omega$ Centauri}
   \author{J. Kaluzny\inst{1}, A. Olech\inst{1}, I.B. Thompson\inst{2}, 
          W. Pych\inst{1}, W. Krzemi\'nski\inst{1,3} and A. Schwarzenberg-
          Czerny\inst{1}
          }

   \offprints{J. Kaluzny}

   \institute{Nicolaus Copernicus Astronomical Center, ul. Bartycka 18,
        00-716 Warszawa, Poland \\
        \email{(jka,olech,pych,alex)@camk.edu.pl}
        \and
        Carnegie Institution of Washington, 813 Santa Barbara Street,
        Pasadena, CA 91101, USA\\
              \email{ian@ociw.edu}
        \and
         Las Campanas Observatory, Casilla 601, La Serena, Chile\\
              \email{wojtek@lco.cl}
             }

   \date{Received ........................., 2004; 
         accepted ........................., 2004}

   \abstract{We present results of a photometric survey for variable
stars in the field of the globular cluster $\omega$ Centauri.  The
observed  region was centered roughly on the cluster core and covered
644 arcmin$^2$. The cluster was monitored on 59 nights  in 1999 and
2000. A total of 117 new variables were identified.  Among them there
are 16 RR Lyr-type stars, 35 SX Phe variables and 26  eclipsing
binaries.\\
A comprehensive catalog including all variable stars so far reported 
from the cluster field is presented. We list basic photometric
properties and provide finding charts for a total of 392 objects.  For
313 of them new $BV$ light curves were obtained.\\
The presented sample includes several interesting variables, such as
SX~Phe stars with extremely short periods of pulsation and several 
candidates for pulsating K giants. Optical counterparts to 9 X-ray sources
detected by  XMM and Chandra telescopes were identified: all of them
are likely to be foreground variables not related to the cluster. 

}

\titlerunning{CASE. Catalog of Variable Stars in
          $\omega$ Cen}
\authorrunning{Kaluzny et al.}
   \maketitle
%

\section{Introduction}

The Cluster AgeS Experiment (CASE) is a long term project aiming at 
determination of accurate ages and distances of nearby globular clusters
(GC) by using observations of detached eclipsing binaries (Paczy{\'n}ski
1997). The project consists of two parts. The first part is an extensive
photometric survey of about 10 Galactic GCs with the aim of identifying 
eclipsing binaries (EB) located near or below the main-sequence turnoff
(MSTO). The survey is conducted on the 1.0-m Swope telescope at Las
Campanas  Observatory. The second part of the project is devoted to
determination of absolute parameters (masses, radii, ages and
luminosities) of selected  EBs. It includes derivation of precise radial
velocity curves  as well as photometric follow up observations in the
optical and near IR domain. The spectroscopic data are currently being
collected, mainly using the 6.5-m Clay and Baade telescopes at Las
Campanas Observatory.  The only system for which we have published  a
complete analysis so far is OGLEGC17=V212 in $\omega$~Cen (Thompson et
al. 2001; Kaluzny et al 2002).

$\omega$~Cen~=~NGC~5139~=~$C1323-472$  [$l=309.10^\circ$, $b=+14.97^\circ$,
$E(B-V)=0.12$; Harris 1996] is the most  luminous and most massive of
all known Galactic GC. An early survey  of that cluster conducted by
Niss et al. (1978) led to identification of the eclipsing binary NJL-5,
which was in fact the first  variable of this type discovered in any
globular cluster. Several EBs were subsequently located in $\omega$~Cen
as results of  a side project conducted by the OGLE-I team (Kaluzny et
al. 1996; Kaluzny et al. 1997a; Kaluzny et al. 1997b). In this
contribution we present the results of an extensive photometric search
aimed at detection of additional variables in the cluster field.
 
\section{Observations}

In the interval from 1999 February 6/7 to 2000 August 9/10 we carried
out CCD photometry on the 1.0-m Swope telescope at Las Campanas
Observatory. The data were collected on 59 nights.  The telescope was
equipped with the SITe3 2k$\times$4k CCD camera with an effective field
of view  $14.5\times 23$ arcmin$^2$ (2048$\times$3150 pixels were used),
at a scale  of 0.435 arcsec/pixel. Two partly overlapping fields, East
and West, covering the central part of the cluster were monitored.  The
full size of the surveyed area was equal to 644 arcmin$^2$.

For the East field we obtained 761 and 199  exposures in the $V$ and $B$
filters, respectively. For the West field we  obtained 594 and 204
exposures in the $V$ and $B$ filters, respectively. Exposure times were
90 s to 300 s for the $V$ filter and from 150 s to 350 s for the $B$
filter, depending on the atmospheric transparency and seeing
conditions.  Median values of the exposure time were 140 s and 200 s for
$V$ and $B$ filters, respectively. The median value of the seeing in the
analyzed frames was 1.2 arcsec.

In addition to the observations listed above we have also obtained for
each of two monitored fields a number of short exposure frames.  These
data, aimed at search for relatively bright variables, will be 
discussed in a separate contribution.

\section{Data reduction}

\subsection{Initial reductions}

Preliminary processing  of the raw CCD frames  was performed using tasks
from the IRAF $ccdproc$  package.\footnote{{\sc iraf} is distributed by
the National Optical Astronomy Observatories, which is operated by the
Association of Universities for Research in Astronomy, Inc., under a
cooperative agreement with the National Science Foundation.}   At that
stage we also used the IRAF $ctio/irlincor$ task to correct for a slight
non-linearity of the CCD camera. 

To search for variables and to obtain their light curves  we employed
the image subtraction package ISIS V2.1 (Alard and Lupton 1998, Alard
2000). As demonstrated by Olech et al. (1999a)  ISIS is capable of
detecting more  variables in the central regions of GCs than surveys
based on profile photometry software such as DAOPHOT or DOPHOT. In our
analysis  we have followed a procedure described in some detail in
Mochejska et al. (2002). Frames with seeing of about 0.9 arcsec taken
during ,,dark time" were selected as reference images. Template images
were constructed for every field and filter combination  by stacking 20
good quality frames which were first geometrically  transformed to the
common grid defined by an appropriate reference image. Before performing
image subtraction each template image was divided into four slightly
overlapping sub-images. All analyzed frames were divided the same way.
This procedure helped minimize effects introduced by variability of the
point spread function across a given image. The sub-images were remapped
to the template sub-frame coordinate system using a second-order
polynomial transformation ({\tt degree = 2}). During this step,
initial rejection of cosmic rays was also performed ({\tt cosmic\_thresh
= 1.0}). Differential brightness variations of the background were
fitted with a second order polynomial transform ({\tt deg\_bg = 1}). A
convolution kernel varying quadratically with position was used ({\tt
deg\_spatial=2}). 

\subsection{Search for variables}

To search for variables we used {\sc find} and {\sc detect} procedures
included in the ISIS package. In particular, {\sc detect}  constructs
two diagnostic images for a list of residual images resulting from the
subtraction procedure. The {\tt VAR} image is the median of the residual
images and  {\tt ABS} is the median of the absolute values of the
residual images. {\sc find} uses these two diagnostic images and returns
a list of potential variables and their positions on the template image.
To increase the chance of finding low ,,duty cycle" variables we divided
our set of frames into 20 subsets which were then analyzed separately.
\footnote{By ,,duty cycle" we mean the fraction of time in which a given
star shows luminosity other than its median luminosity. Examples of 
stars with high and low ,,duty cycles" are RR Lyr stars and detached
eclipsing binaries with narrow eclipses, respectively.}  Each subset
contained images from 2-4 consecutive nights. A separate search was
performed for the full list of available images.  The list of candidate
variables was increased by adding all stars lying within the blue
straggler region of the cluster color-magnitude diagram. This was done
in order to identify low amplitude SX Phe stars and contact binaries
which are quite common among blue stragglers.

For each potential variable, its light curve expressed in {\tt ADU} 
units was extracted using the {\sc phot} task of the ISIS package.  The
search for periodic signals in these light curves was performed  using a
program  based on the {\sc anova} statistic introduced by  
Schwarzenberg-Czerny (1996). Each light curve, phased with the most
probable period, was then examined  by eye.

\subsection{Conversion of light curves to instrumental magnitudes}

The ISIS package allows extraction of  light curves expressed in
differential  counts. Conversion of a light curve for a given object to
magnitude units is possible provided that one is able to derive the
total flux  of that star on the template image. This can be accomplished
by  deriving profile photometry and an appropriate aperture correction
for a given field. We used the DAOPHOT/ALLSTAR package (Stetson 1987)
along with the  DAOGROW program (Stetson 1990) to perform this job. 
Some of the variables identified with the ISIS software were located in
the highly crowded region near the cluster core or in the vicinity of
some saturated stars. For such  objects ALLSTAR was unable to provide
reliable magnitudes and their light curves were left in differential
{\tt ADU} units.

\subsection{Transformation to the standard system}

Transformation from the instrumental magnitudes to the standard $BV$
system was established based on observations of several Landolt (1992)
fields obtained in 1999 and 2000 seasons. It was found that the 
instrumental system was very stable  over two observing seasons.
Therefore the color terms of the transformation were derived by
averaging results obtained on several photometric nights.  The following
relations were  adopted:

\begin{eqnarray}
v=V-0.014\times (B-V) +0.110\times X +2.927\\
b=B-0.043\times(B-V)+0.202\times X +3.140
\end{eqnarray}

\noindent where $X$ is the airmass and lower case letters correspond
to the instrumental magnitudes. The extinction coefficients were
derived based on the data collected  on the night of 1999 Jun  19/20.
Twenty nine observations of 26 stars from 5 Landolt fields were obtained
on that night\footnote{One of the fields was observed twice. Landolt
list of standards  was extended by adding 4 objects from  Stetson
(2000).}.  The standard fields  were observed at air-masses spanning
the range 1.07-1.71. In Fig. 1 we show the residuals of the standard
stars resulting from the above transformations. 

\subsection{Astrometry}

Transformation from rectangular frame coordinates to equatorial 
coordinates was derived based on positions of a set of bright stars
selected from the proper motion study published by van Leeuwen et al.
(2000). We used 1976 and 1320 transformation stars for fields W and E,
respectively. The adopted frame solutions reproduce the equatorial
coordinates of Leeuwen et al. (2000) with residuals rarely exceeding
0.6 arcsec.

\section{The catalog}

\subsection{Basic properties of the variables}

Clement et al. (2001) list 293 variable stars in the field of $\omega$
Centauri. Stars V28, V31, V37 and V93, after work of Bailey 
(1902), are marked as "not variable" -- our photometry confirms that
finding. We found that also the objects V176, V187, V188, V189, V190,
V191, V193, V278, V279 and V290 show no signs of variability.
Additionally, some of variables listed by Clement et al. (2001) are
included twice in their catalog (V256 is the same as  V231, V262=V216,
V286=V94 and V287=V169). There are a total of 275 genuine variables in
the field of  $\omega$ Centauri which are listed in the paper of
Clement et al. (2001). Sixty of them were located outside the field
covered by our survey. Additionally, images of 19 bright variables were
severely saturated in our images and as result no photometry could be
obtained for these  objects. We present $BV$ light curves for a total of
196 out of  275 objects included in the catalog of Clement et al.
(2001).  We also report identification and provide $BV$ light curves for
117  newly identified variables. Among them are 17 RR Lyr-type stars, 
35 SX Phe variables and 26 eclipsing binaries.
 
Table 1 presents some basic data for a total of 392 variables 
identified so far in the field of $\omega$ Centauri\footnote{The full
version of Table 1 is only available at the CDS}. The columns  of this
table are as follows:

\begin{itemize}

\item {\it Star}: ~Star number. Numbering scheme of variables V1--V293
is taken from Clement et al. (2001). Objects designated as "NV" are new
discoveries.

\item {\it RA}: ~Right ascension of the star for epoch J2000.0

\item {\it Decl.}: ~Declination of the star for epoch J2000.0.

\item {\it Period}: ~Period of variability in days.  For the
majority of previously known variables, this quantity is taken from
Martin (1938), Sawyer Hogg (1973) or Kaluzny et al. (1996, 1997a,b). In
cases where their period does not fit the data, we provide a new
determination. For variables NV294--NV410 the period was derived using
the {\sc anova} statistics of Schwarzenberg-Czerny (1996).

\item $V_{max}$: ~Maximal brightness in the $V$-band.\footnote{
For stars which are located outside of the OGLE and CASE
fields, the $V$ and $B$ (photographic) magnitudes are taken
from Dickens and Carey (1967) or Geyer and Szeidl (1970).
For the remaining variables we present the maximum and
minimum photographic ($m_{pg}$) magnitudes from Martin
(1938) or Sawyer (1955). Since these are blue sensitive,
they are designated as `B' in Table 1.}

\item $V_{min}$: ~Minimal brightness in the $V$-band.$^4$

\item $<V>$: ~Intensity averaged mean $V$ magnitude of the star. This
field is left empty for eclipsing binaries.$^4$

\item $<B>$: ~Intensity averaged mean $B$ magnitude of the star or
maximal brightness in the $B$-band for eclipsing binaries.

\item {\it Type}: ~Type of variability. Below we list the meanings of
the symbols used:

\begin{itemize}

\item RR0: fundamental mode RR Lyr star (RRab) 

\item RR1: 1st overtone RR Lyr star (RRc). In fact, all RR Lyr
stars which were not classified as fundamental mode pulsators were
designated as RR1. Clement and Rowe (2000) identified as many as 21
possible second overtone pulsators (RR2) in $\omega$ Cen but detailed
analysis of the light curves of these stars indicates that they are
mostly RR1 stars with strong non-radial peaks (Olech et al., in
preparation)

\item SX: SX Phe type variable

\item EA: detached eclipsing binary, out of eclipses light curve is flat

\item EB: eclipsing binary,  out of eclipses light curve is not flat

\item EW: contact binary (W UMa type variable)

\item C: population II Cepheid

\item ell: candidate ellipsoidal variable

\item irr: irregular light curve

\item SR: semi-regular variable

\item sp: candidate spotted variable

\item BL: BL Her type star

\item LT: long-term irregular or long-period variable

\item CV: candidate cataclysmic variable

\item RV: RV Tauri type star

\end{itemize}

\item {\it Other design.}: ~Designations used in some previous studies of
variable stars in $\omega$ Centauri (e.g. Kaluzny et al. 1996, 1997a, 1997b;
Niss et al. 1978).

\item {\it Remarks}: ~General remarks, where the following symbols mean:

\begin{itemize}

\item pc: ~period change

\item mp: ~multi-periodic star

\item Bl: ~Blazhko effect

\item out: ~star located out of our field

\item non var: ~no traces of variability in our data

\item saturated:~stellar image saturated in one or two bands

\end{itemize}

\end{itemize}

\begin{table*}[!t]
\caption{Basic properties of variable stars in $\omega$ Centauri}
{\small
\begin{tabular}{|l|c|c|r|c|c|c|c|l|l|l|}
\hline
\hline
Star & RA [2000] & Decl. [2000] & Period & $V_{max}$ & $V_{min}$ & $<V>$ & $<B>$ & Type & Other & Remarks\\
     & $~~^h~~^m~~^s$ & $~~~^\circ ~~~'~~~"$ & [days] & & & & & & design. & \\
\hline
V1 & 13 26 05.26 & $-$47 23 42.8 & 29.3479 & 10.40 & 11.55 & 10.89 & 11.66 & RV &  & saturated \\
V2 & 13 26 12.65 & $-$47 24 41.9 & 235.74 & 12.5: & 16.61 & 13.825 & 14.631 & LT & GC23 & \\ 
V3 & 13 25 56.15 & $-$47 25 53.8 & 0.841258 & 14.01 & 14.76 & 14.389 & 14.907 & RR0 & GC184 & \\
V4 & 13 26 12.93 & $-$47 24 18.8 & 0.627320 & 13.83 & 14.87 & 14.453 & 14.905 & RR0 & GC99 & \\
V5 & 13 26 18.33 & $-$47 23 12.4 & 0.515274 & 14.0: & 15.15 & 14.745 & 15.235 & RR0 & GC101 & Bl\\
V6 & 13 26 30.24 & $-$47 24 28.4 & 110: & 13.84$B$ & 15.24$B$ & - & 14.363 & LT &  & saturated in $V$\\
V7 & 13 27 00.90 & $-$47 14 00.5 & 0.7130 & 14.08 & 15.01 & 14.59 & - & RR0 & GC87 & out\\ 
V8 & 13 27 48.45 & $-$47 28 20.3 & 0.521329 & 13.90 & 15.16 & 14.683 & 15.061 & RR0 & GC199 & \\
V9 & 13 25 59.58 & $-$47 26 24.0 & 0.523480 & 14.1: & 15.19 & 14.756 & 15.274 & RR0 & GC183 & Bl\\
V10 & 13 26 06.99 & $-$47 24 36.6 & 0.374976 & 14.31 & 14.68 & 14.484 & 14.902 & RR1 & GC98 & \\
\hline
\hline
\multicolumn{11}{l}{B - $B$ magnitudes available only}\\
\end{tabular}}
\end{table*}

\subsection{The finding charts}

The finding charts for all known or suspected variables located in the
field of $\omega$ Centauri are shown in Fig. 2\footnote{Figure 2 is only
available in electronic form at http://www.edpsciences.org}. For
variables placed out of our fields charts were created  using template
frames of Kaluzny et al. (1996, 1997a,b) or DSS\footnote{ The Digitized
Sky Survey was produced at the Space Telescope Science  Institute under
U.S. Government grant NAG W-2166. The images of these  surveys are based
on photographic data obtained using the Oschin Schmidt  Telescope on
Palomar Mountain and the UK Schmidt Telescope. The plates were 
processed into the present compressed digital form with the permission
of  these institutions.} images.  The finding charts based on images
from our survey or from  Kaluzny et al. (1996, 1997a, 1997b)  have size 
$28\arcsec\times28\arcsec$. Charts based on the DSS images have size
$60\arcsec\times 60\arcsec$. On all charts north is up and east is to
the left.
 
\subsection{Comparison with OGLE data}

For a total of  173 variables observed in our survey  photometry was
also published by the OGLE-I group (Kaluzny et al. 1996, 1997a, 1997b).
Differences in mean magnitudes for these variables are plotted in Fig.
3. The mean difference in the sense of \linebreak $<V>_{\rm CASE} -
<V>_{\rm OGLE}$ is $-0.016\pm0.006$ mag. Moreover, one can see that
$<V>_{\rm CASE} - <V>_{\rm OGLE}$ vs. $<V>_{\rm CASE}$ relation  shown
in Fig. 3 is not flat. Fitting straight line to the data from Fig. 3
gives the following relation:

\begin{equation}
<V>_{\rm CASE} - <V>_{\rm OGLE} ~=~ -0.020 <V>_{\rm CASE} + 0.293 
\end{equation}

The mean magnitudes of bright stars ($V\approx14$ mag) are close to each
other in both compared data sets. A substantial  difference of
$\sim0.05$ mag is observed for fainter variables with $V\approx17$. The
likely source  of such a discrepancy is  nonlinearity of the CCD camera
used by the OGLE-I group (Paczy\'nski et al. 1999).

\subsection{Some interesting objects and location of variables on the
CMD}

Phased light curves of 117 newly identified variables are presented
in Figures 4, 5, 6 and 7 which show RR Lyr, SX Phe, eclipsing binaries 
and remaining types of variables, respectively. Note that in some 
cases light curves are displayed in differential counts units. 

Figure 8 shows  positions of RR Lyr and SX Phe variables from Table 1 in
the  color-magnitude diagram of the cluster. The brightest of SX~Phe
star,  V65, is a foreground object (see Sec. 5). We also note that the
color of the formally bluest of the SX Phe stars, variable NV321 , is
very uncertain.

Figure 9 presents the color-magnitude diagram of $\omega $ Cen with 
positions of all objects from Table 1 marked apart from RR Lyr and SX
Phe variables. Eclipsing binaries are denoted with  squares while
triangles denote other classes of variables.  About 50\% of variables
are located in the region occupied by  cluster blue stragglers (BS). In
particular one can notice presence of a group of relatively bright BS
candidates with $16<V<16.6$. Membership status of these stars, as well
as other variables, can be  determined relatively easily by measuring
their radial velocities. We note that the cluster itself has large
radial velocity  $V_{rad}=232.3$~km/s (Harris 1996) and therefore field
interlopers are relatively easy to distinguish from the cluster members.

There are several interesting objects among newly identified  variables.
For example, NV294 and NV295 are the shortest period SX Phe type stars
known. Up to now the shortest period variable of this type was V10 from
NGC6397 (Kaluzny and Thompson 2003). Objects NV361, NV362 and NV363 are
probably main-sequence detached eclipsing binaries. They may turn out to
be excellent targets for follow up spectroscopic study aimed at
determination  of the age and/or distance to the cluster (Thompson et
al. 2001, Kaluzny et al. 2002). RRc type variables NV339, NV340, NV342,
NV351 and NV399 are likely members of the recently identified group of
non-radial pulsators (Olech et al. 1999a, 1999b, 2001, Clement and  Rowe
2000). Variable NV383 shows regular modulations with a period of 8.6
days superimposed with $\sim 0.6$~mag outbursts or flares occurring on a
time scale of tens of days.

Among RR Lyr variables with known period we list 86 RR0 and 100 RR1.
Evidence of period change was found in 14 RR1 and one RR0. For
these variables it was not possible to fit one period to all of the
data. Often this change amounts to as little as $2\sigma$. However, its
correlation with the RR1 type is impressive: the probability of getting
at least 14 RR1 stars by random drawing 15 stars from our sample is as
little as $(99/186)^{14}=0.00015$. A detailed analysis of RR Lyr type
variables in the fields of $\omega$ Cen will be given in a forthcoming
paper (Olech et al., in preparation).

Two variables, NV380 and NV404, are located among stars  forming an
extreme horizontal branch (equivalent to population of  field sdB star) 
in the color-magnitude diagram. They show low amplitude sine-like light 
curves with periods of 7.8~d and 5.2~d for NV380 and NV404,
respectively.  It is well established that sdB stars are formed in
binary systems (eg. Morales-Rueda et al. 2003) and therefore it is
likely that  observed variability of NV380 and NV404 is related to their
binarity.  

In Fig. 9 one may note the presence of several variables distributed
along the red giant branch of the cluster. About a dozen of these
variables show low amplitude, periodic light curves with $\delta V < 
0.1$ and $P < 10$~d.   At least some of these stars are likely analogues
of variable K giants identified in 47~Tuc by Edmonds \& Gilliland
(1996). It has been speculated  that the underlying source of
variability of 47 Tuc K giants are low-overtone  pulsations.
Particularly interesting is NV397 whose light curve is modulated
with an extremely short, as for a giant star, period of $P=0.273$~d.

\section{X-ray counterparts}

Omega Centauri is known to posses a rich population of X-ray sources.
Recently, the Chandra telescope observed $17\times 17$~arcmin$^{2}$
field centered approximately on the cluster core.  Cool et al. (2002)
reported detection of over 140 sources in that field. Unfortunately they
have so far not published coordinates of these objects, which makes the
search for optical counterparts difficult. Rutledge et al. (2002)
independently analyzed Chandra observations and published accurate
coordinates of 37 X-ray sources from the cluster field. Gendre at al.
(2003) reported identification and coordinates  of 146 X-ray sources
inside a $23\times 23$~arcmin$^{2}$ field observed by the XMM-Newton
telescope. Position errors for these sources, as derived from XMM
images, range from a fraction of arcsec up to  over 40 arcsec with a
median value of 5.4 arcsec. As noted by the authors an additional source
of uncertainty is the systematic error of the pointing direction of
satellite which  is estimated at 4 arcsec. For 64 XMM sources Gendre at
al. (2003) were able to identify their Chandra counterparts and
consequently provide accurate coordinates with uncertainty of the order
of 1 arcsec.

In Table 2 we list candidates for optical counterparts of X-ray sources
which were selected by cross-correlation of our catalog of variables
with objects from Gendre at al. (2003; their Table 3) and Rutledge
et al. (2002; their Table 1). All optical variables with counterparts
identified among sources from  Gendre at al. (2003) have also counterparts
among objects listed in Rutledge et al. (2002).

The objects corresponding  variables  NV367=XMM-3 and NV375=XMM-7 were
earlier reported by Cool et al. (1995) as optical counterparts of
Einstein-HRI  sources D and A, respectively\footnote{ Einstein sources D
and A correspond to ROSAT sources 4 and 3, respectively (Verbunt \&
Johnston 2000).}. Cool et al. (1995) obtained optical spectra of both
stars and concluded  that they were foreground dM2e-dM3e dwarfs. We note
that  NV367 and NV375 show rather stable sine-like light curves with a
peak-to-peak range of about 0.1 mag in the $V$-band. Stability of the
light curves on a time scale of one year points toward binarity as a
cause of the observed variability.

Variable N378=XMM-31 is a detached eclipsing binary whose location on
the  cluster color-magnitude diagram indicates that it is a foreground
object.

The optical light curve of star NV383=XMM-95 shows modulation with
P=8.63~d with superimposed occasional ,,flares" lasting a few hours and
having amplitudes  of  a few tenths of magnitude. This variable, as well
as the remaining stars included in Table 2, are most likely field
objects not related to the cluster. This conclusion is based on the
observed location of these stars on the cluster color-magnitude diagram.
It is also in agreement with the proper motion study of van Leeuwen
et. al (2000). Variables V167, V223, NV367, NV375 and NV378, which are
bright enough to be included in that study, have probability of
membership in range of 0--2\% and most probably are not cluster members.

\setcounter{table}{1}
\begin{table}[h]
\caption{The XMM sources correlated with CASE variables. Last column
gives difference between optical and X-ray position.}
{\small
\begin{tabular}{rrr}
\hline
\hline
XMM & CASE & $\Delta$ \\
    &      &[arcsec]   \\
\hline
 3  & 367 &   1.7    \\
 7  & 375 &   1.3    \\
 25 & 167 &   1.0    \\
 29 & 223 &   2.1    \\
 31 & 378 &   1.2    \\
 33 & 376 &   3.1    \\
 68 & 377 &   0.4    \\
 74 & 369 &   1.5    \\
 95 & 383 &   2.5    \\
\hline
\hline
\end{tabular}}
\end{table}

\section{Summary}

An extended survey of the central part of $\omega$~Cen led to identification
of 117 new variables. We present $BV$ light curves for 110 of them along
with light curves for 196 previously known variables.

Photometry presented in this paper will be discussed in some detail in
separate contributions devoted to specific classes of variables. A
paper aimed at calibration of absolute magnitude $M_{V}$ for RRab and
RRc stars has already been published (Olech et al. 2003) and the paper
about SX Phe-type variables in $\omega$ Cen is in preparation.

We are in a process of obtaining spectroscopic data suitable for
checking the membership status of detached EBs identified to date in the
cluster field.

The electronic version of this catalog is available via Internet at
the following home page of the CASE project: {\tt http://case.camk.edu.pl}

\begin{acknowledgements}

ASC, JK, WK and WP were supported by KBN grant  5~P03D004.21. AO was
supported by KBN grant 2~P03D024.22. We are grateful to the referee Dr.
Christine Clement for her helpful comments and suggestions.

\end{acknowledgements}

\noindent{\bf Appendix A: Notes on individual variable stars}

\begin{description}

\item {\bf V1} Photographic $B$ and $V$ data listed are from Dickens and
      Carey (1967).

\item {\bf V6}: Saturated in the $V$-band. Light curve transformed 
      into magnitudes only for the $B$-band. Photographic
      photometry listed after Martin (1938).

\item {\bf V7}: Out of surveyed field. Finding chart and photometric data 
      based on  Kaluzny et al. (1997b)

\item {\bf V17}: Light curve transformed into instrumental magnitudes
only for the $B$ band. The star is saturated in our images
for the V band. The photometric data listed are $B$ data
from Martin (1938). Sawyer Hogg (1973) classified this star
as an irregular and listed a period of 64.725 days, but the
64 day period does not fit our data for the $B$ band.

\item {\bf V22}: Clear change of period from 0.39606(3) to 0.39624(6)
      days between 1999 and 2000 seasons.

\item {\bf V23}: Variable present in both surveyed sub-fields.

\item {\bf V29} Photographic $B$ and $V$ data listed are from Dickens and
      Carey (1967).

\item {\bf V30}: Clear change of period from 0.40441(6) to
      0.40410(10) days between 1999 and 2000 seasons.

\item {\bf V36}: Out of surveyed field. Finding chart and photometry 
      based on  Kaluzny et al. (1996).

\item {\bf V48}: Saturated in $V$. Light curve transformed into magnitudes
      only for the $B$-band.  Photographic $B$ and $V$ data listed are from 
      Dickens and Carey (1967).

\item {\bf V52}: Jurcsik et al. (2001) classified this star as
"above horizontal branch" because it is significantly brighter
than the RR Lyrae variables.
Our data indicate that it is RRab star. According
      to van Leeuwen et al. (2000) the probability of membership for this
      variable is only 45\% and most probably it is a foreground
      star. Its low amplitude, compared with other stars of the same
      period, may suggest high metallicity.

\item {\bf V55}: Out of surveyed field. Finding chart based on DSS. 
      Photometric data listed after Geyer and Szeidl (1970).

\item {\bf V58}: Change of period from 0.36988(2) to 0.36975(6) days
      between 1999 and  2000 seasons.

\item {\bf V63}: Out of surveyed field. Finding chart based on DSS. 
      Photometric data listed after Geyer and Szeidl (1970).

\item {\bf V65}: This variable corresponds to the star number 60026 from
      the proper motion study of van Leeuwen et al. (2000). The 
      probability of membership for this variable is 0\%, thus is 
      must be a foreground star.

\item {\bf V66}: Change of period from 0.40727(3) to
      0.40761(9) days between  1999 and 2000 seasons.

\item {\bf V68}: Change of period or phase shift between 1999 
      and 2000 seasons. The period changed from 0.53461(2) to
      0.53509(9) days

\item {\bf V69}: Out of surveyed field. Finding chart based on  DSS. 
       Photometric data listed after Geyer and Szeidl (1970).

\item {\bf V72-V73}: Out of surveyed field. Finding chart based on  DSS.
       Photometric data listed after Geyer and Szeidl (1970).

\item {\bf V79}: Out of surveyed field. Finding chart and photometric data 
      based on  Kaluzny et al. (1996).

\item {\bf V80}: Out of surveyed field. Finding chart based on  DSS.
       Photometric data listed after Sawyer (1955).

\item {\bf V84-V85}: Out of surveyed field. Finding chart and photometric 
      data based on  Kaluzny et al. (1996). 

\item {\bf V89}: Change of period from 0.37511(7) to 
      0.37475(13) days between  1999 and 2000 seasons. 
      Variable present on both surveyed  sub-fields.

\item {\bf V90}: Variable present on both surveyed  sub-fields.

\item {\bf V91}: Variable present on both surveyed  sub-fields.

\item {\bf 92}: Based on observations made in the early 1930s,
Martin (1938) derived a period of 1.3450659 days.
Our data for the 1999-2000 seasons indicate that the
period is significantly longer, 1.34604(34) days, in
agreement with the period change rate, $13.941\pm0.513$ days
per Myr derived by Jurcsik et al. (2001)

\item {\bf V94}: Our $x$ coordinate for this star ($-479.5$)
differs from the value ($-504.09$) published by Martin (1938). This
star is the same as V286. Multi-periodic.

\item {\bf V110}: Multi-periodic. The period of 0.3221021 days listed 
      by Martin (1938) does not fit our data.

\item {\bf V111}: Variable present in both  surveyed sub-fields.

\item {\bf V114}: Variable present in both  surveyed sub-fields.

\item {\bf V129}: Light curve transformed into instrumental
magnitudes only for the $B$ band. The star is saturated in our images
for the $V$ band.

\item {\bf V131}: Clear change of period from 0.39225(5) to
      0.39198(8) days between 1999 and 2000 seasons.

\item {\bf V133}: Out of surveyed field. Finding chart based on DSS. 
      Photographic magnitudes given after Martin (1938).

\item {\bf V134}: Out of surveyed field. Finding chart based on DSS.
      Photometric data given after Geyer and Szeidl (1970).

\item {\bf V135}: The light curve is noisy. Possibly multi-periodic star.

\item {\bf V140}: Sawyer (1955) indicated that this variable has a
short period, but did not determine a value.
Our data indicate that it is RRab star showing very strong Blazhko effect.

\item {\bf V142}: Sawyer (1955) indicated that this variable has a
short period, but did not determine a value.
Our data indicate that it is RRc star showing also pulsations in 
non-radial modes.

\item {\bf V144}: Variable present in both surveyed sub-fields.

\item {\bf V149}: Out of surveyed field. Finding chart based on  DSS. 
      Photometric data listed after Geyer and Szeidl (1970).

\item {\bf V150}: Based on observations made in the early 1930s,
Martin (1938) derived a period of 0.8991585 days.
Our data for the 1999-2000 seasons indicate a longer
period, 0.89930(3) days, in agreement with the period
change rate, $2.476\pm0.67$ days per Myr derived by
Jurcsik et al. (2001).

\item {\bf V151}: Out of surveyed field. Finding chart based on  DSS. 
      Photometric data listed after Martin (1938).

\item {\bf V153}: Martin (1938) listed period of 0.3864509 days. 
      Our data indicate for 1999-2000 seasons the period of 0.38624(2) days.

\item {\bf V154}: Martin (1938) listed period of 0.3223311 days. 
      Our data indicate the period of 0.32234(2) days for 1999-2000 seasons.
      Variable present in both surveyed sub-fields.

\item {\bf V156}: Change of period from 
      0.35913(2) to 0.35901(2) days between 1999 and 2000 seasons. Variable 
      present in both surveyed sub-fields.

\item {\bf V157}: Phase shift between 1999 and 2000 seasons. Variable present 
      in both surveyed sub-fields.

\item {\bf V158}: Martin (1938) listed period of  0.367335 days. 
      Our data indicate for 1999-2000 seasons the period of 0.367276(3) days.
      Variable present in both surveyed sub-fields.

\item {\bf V159-V160}: Out of surveyed field. Finding charts based on DSS. 
      Photometric data listed after Geyer and Szeidl (1970).

\item {\bf V166}: Multi-periodic. Variable present in both surveyed sub-fields.

\item {\bf V170}: Out of surveyed field. Finding chart based on DSS.

\item {\bf V171-V175}: Out of surveyed field. Finding charts based on DSS.

\item {\bf V176}: There is a bright star at the position of V176 but in 
      our data it shows no variability. In the catalog of Sawyer Hogg (1973) 
      this star is classified as RRc variable.

\item {\bf V177-V183}: Out of surveyed field. Finding charts based on DSS.

\item {\bf V187-V191}: Variables classified as possible eclipsing 
      binaries by Niss, Jorgensen \& Laustsen (1978). No signs of 
      variability in our data.

\item {\bf V193}: Variable classified as possible eclipsing binary by Niss,
      Jorgensen \& Laustsen (1978). No signs of variability in our data.

\item {\bf V198}: Large error of the zero point for $V$ curve (expected error 
      of about 0.1 mag).

\item {\bf V210}: Variable present in both surveyed sub-fields.

\item {\bf V211}: Variable present in both surveyed sub-fields.

\item {\bf V216}: This variable is the same as V262 (GC22=GC95).
The highest peak in the power spectrum of the light curve is found at
the frequency corresponding to the period of 23.737 days. There is only
marginal peak around period of 0.4881 days reported by Kaluzny et al.
(1997b), thus we conclude that the true period is 23.737 days.

\item {\bf V220}: Variable present in both surveyed sub-fields.

\item {\bf V229}: $V$ magnitude almost the same as in the OGLE
      data. Our observations show the star significantly bluer
      than in OGLE data. Something wrong with OGLE $I$ magnitude
      or our $B$ magnitude determination.

\item {\bf V231}: This variable is the same as V256 (GC66=GC38)

\item {\bf V233}: Variable present in both surveyed sub-fields.

\item {\bf V239}: Strong blending. Magnitudes and colors
                   uncertain.

\item {\bf V242}: Variable classified as eclipsing binary by 
      Kaluzny et al. (1997a). There are no eclipses in our data.

\item {\bf V251}: Period of 0.92247(15) days fits our data better than 
      the period of 0.63093 days given by Kaluzny et al. (1997a)

\item {\bf V256}: This variable is the same as V231 (GC66=GC38)

\item {\bf V262}: This variable is the same as V216 (GC22=GC95)

\item {\bf V263}: Variable classified as RRc star with period 0.5028 days 
      by Kaluzny et
      al. (1997b). Our data indicate that it is RRab or C 
      star with a period of 1.012158 days. 

\item {\bf V265}: Change of period from 0.42255(5) to 0.42309(7) days
      between 1999 and 2000 seasons.

\item {\bf V269}: Variable classified as C or BL Her star with
      period close to one day by Kaluzny et al. (1997b). Our data indicate
      that it is a long term variable with a period of about 145 days.

\item {\bf V270}: The period given by Kaluzny et al. (1997b) is 0.2381 days. 
      Our data indicate period of 0.312714(3) days. That period fits also
      data of Kaluzny et al. (1997b). We conclude that the longer period 
      is real one and thus V270 is an ordinary RRc star and not second 
      overtone RR Lyr star as was proposed by Clement and Rowe (2000).

\item {\bf V271}: Change of period from 0.44298(5) to
      0.44339(9) days between 1999 and 2000 seasons. Variable present in both 
      surveyed sub-fields.

\item {\bf V274}: Variable present  in both surveyed sub-fields.

\item {\bf V275}: Multi-periodic. Variable present in both surveyed sub-fields.

\item {\bf V278-V279}: Variables classified as possible C 
      by Kaluzny et al. (1997b). Both stars had periods very close to one 
      day and had noisy light curves. No signs of variability in our data.

\item {\bf V286}: This variable is the same as V94.

\item {\bf V287}: This variable is the same as V169

\item {\bf V290}: Variable classified as possible RRab star by Kaluzny et
      al. (1997b). No signs of variability in our data.

\item {\bf V292}: Variable number 50259 in van Leeuwen et al.
(2000).

\item {\bf V293}: Variable number 55071 in van Leeuwen et al.
(2000).

\item {\bf NV294}: Very low amplitude SX Phe variable. It is placed in the blue
       straggler region of the CMD.

\item {\bf NV295}: Very low amplitude SX Phe variable. It is placed in the blue
       straggler region of the CMD. Other possible period is 0.0185707
       days. 

\item {\bf NV296}: Very low amplitude SX Phe variable. It is placed in the blue
       straggler region of the CMD.

\item {\bf NV297}: Very low amplitude SX Phe variable. It is placed in the blue
       straggler region of the CMD. Power spectrum indicates presence of
       at least two distinct frequencies in the light curve.

\item {\bf NV302}: Other possible value of the main  period is 0.035516 days.
       Probably multi-periodic star.

\item {\bf NV303}: Other possible value of the main period is 0.035950 days.  
       Probably multi-periodic star.

\item {\bf NV311}: Variable placed in the crowded region. 
      Light curve left in differential ADU counts.

\item {\bf NV312}: Other possible value of the main period is 0.041527 days.

\item {\bf NV313}: Fundamental mode pulsator.

\item {\bf NV314}: Three bright stars near variable found. Zero point
      in $V$-band determined with uncertainty of 0.05 mag.

\item {\bf NV315}: Multi-periodic. Variable present in both surveyed sub-fields.

\item {\bf NV317}: Multi-periodic. Other possible period is 0.0427515 days.

\item {\bf NV320}: Variable present  in both surveyed sub-fields.

\item {\bf NV321}: Variable present in both surveyed sub-fields 
       Magnitudes and color uncertain due to strong blending.

\item {\bf NV324}: Fundamental mode pulsator.

\item {\bf NV326}: A classic double mode star.

\item {\bf NV327}: Variable present  in both surveyed sub-fields.

\item {\bf NV329}: Low amplitude contact binary or ellipsoidal variable.
       Variable placed in the crowded region. Light curve left in 
       differential ADU counts.

\item {\bf NV330}: Variable placed in the crowded region. 
      Light curve left in differential ADU counts.

\item {\bf NV333}: Faint variable placed in the crowded region. 
      Light curve left in differential ADU counts.

\item {\bf NV337}: Very strong blending in $B$. Difficulties in
      deriving the $B$-band light curve. It may also be a sdB star.

\item {\bf NV343}: Variable present  in both surveyed sub-fields.

\item {\bf NV346}: Variable present  in both surveyed sub-fields.

\item {\bf NV347}: Variable present  in both surveyed sub-fields.

\item {\bf NV348}: The light curve in $V$-band is noisy. $B$-band light 
       curve is more stable. Variable present  in both surveyed sub-fields.

\item {\bf NV349}: Due to the vicinity of bright companion star we were 
       only able to obtain a light curve in the $B$-band and only in 
       differential ADU counts.

\item {\bf NV351}: Multi-periodic. Variable placed in the crowded region. 
       Light curve left in differential ADU counts.

\item {\bf NV352}: The star located in the vicinity of bright and 
       saturated star.
       The magnitude and color determinations may be uncertain.

\item {\bf NV353}: Multi-periodic. Variable identified in both surveyed 
       sub-fields. Poor photometry for East sub-field due to the position near 
       edge of the frames.

\item {\bf NV356}: Variable placed in the crowded region. 
      Light curve left in differential ADU counts.

\item {\bf NV366}: Variable with period very close to 1 day. Our 
      observations cover the light curve only near maximum. 
      The formally derived mean magnitude of the star is therefore 
      overestimated.

\item {\bf NV370}: Other possible period is 1.78786 days.

\item {\bf NV372}: Other possible period is 2.6302 days.

\item {\bf NV373}: Other possible period is 1.4571 days.

\item {\bf NV378}: Eclipsing binary with light curve likely affected
                   by chromospheric  activity.

\item {\bf NV379}: Other possible period is 7.225 days.

\item {\bf NV383}: Periodic variable with quasi-periodic 
                   outbursts.

\item {\bf NV387}: Two bright stars in the vicinity of the variable.
                   Difficulties in deriving $B$-band light curve.

\item {\bf NV389}: The light curve can be phased with an alternative period
                   of 30.39~d. In such a case the variable would be classified
                   as an eclipsing binary.

\item {\bf NV391-392}: Variable saturated in the $V$-band. Only $B$-band 
                   light curve available. 

\item {\bf NV394}: Variable saturated in the $V$-band. Only $B$-band 
                   light curve available.

\item {\bf NV395}: Variable located in a crowded region. $V$-band light 
                   curve left in differential ADU counts.

\item {\bf NV403}: Suspected variable with quasi-periodic or 
                   non-periodic outbursts. $P=3.9833$ d in 1999.

\item {\bf NV404}: The light curve can be phased with an alternative period 
                   of 0.8359~d.

\item {\bf NV405}: Suspected variable. Irregular ,,flares"?

\item {\bf NV406}: Variable star showing rapid and deep dips in the 
                   light curve.

\item {\bf NV407}: Possible eclipsing binary star with only one eclipse 
                   present in our data.

\item {\bf NV408}: Variable star showing one distinct outburst in our data.

\item {\bf NV409}: Possible eclipsing binary star.

\item {\bf NV410}: The light curve can be phased with an alternative period 
                   of 20.165~d.  
\end{description}

\clearpage

   \begin{figure*}[!h]
   \centering
\includegraphics[scale=.87]{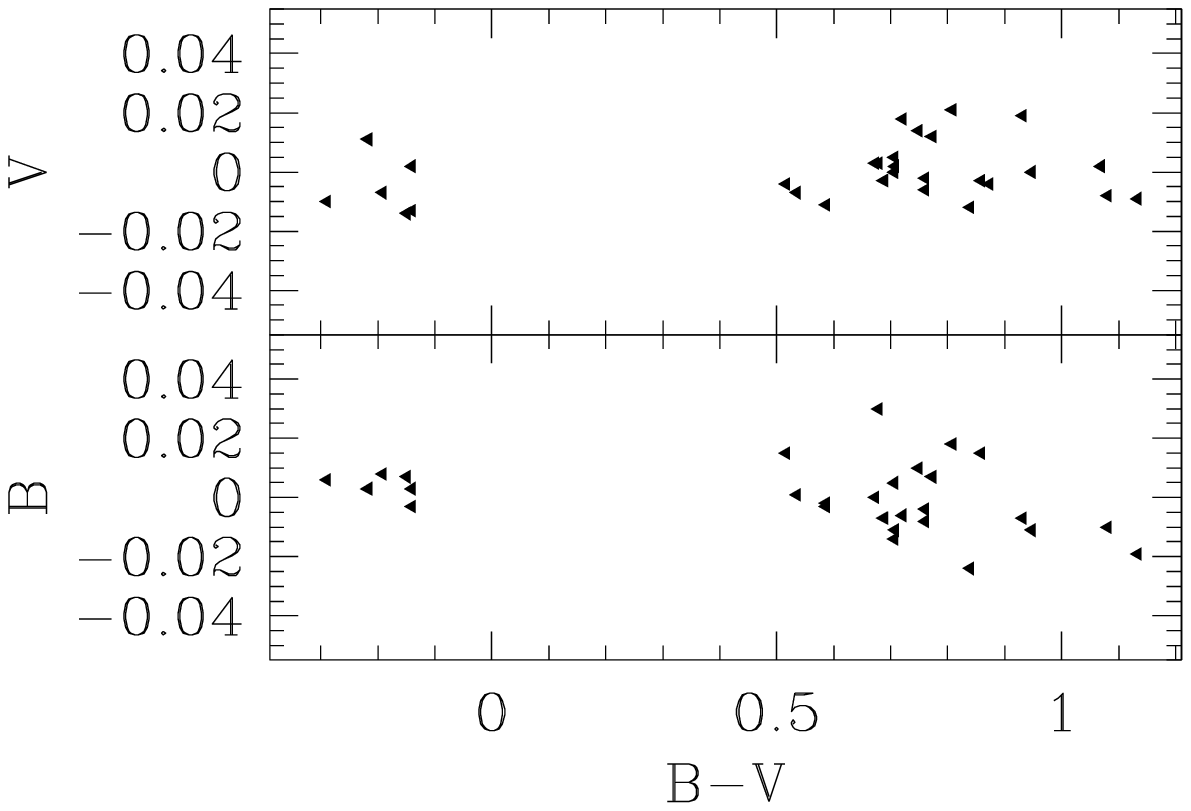}
      \caption{Residuals for Landolt standards observed on the night
of 1999 June 19/20
              }
         \label{FigVibStab}
   \end{figure*}

\clearpage

   \begin{figure*}[!h]
   \centering
\includegraphics[scale=.90]{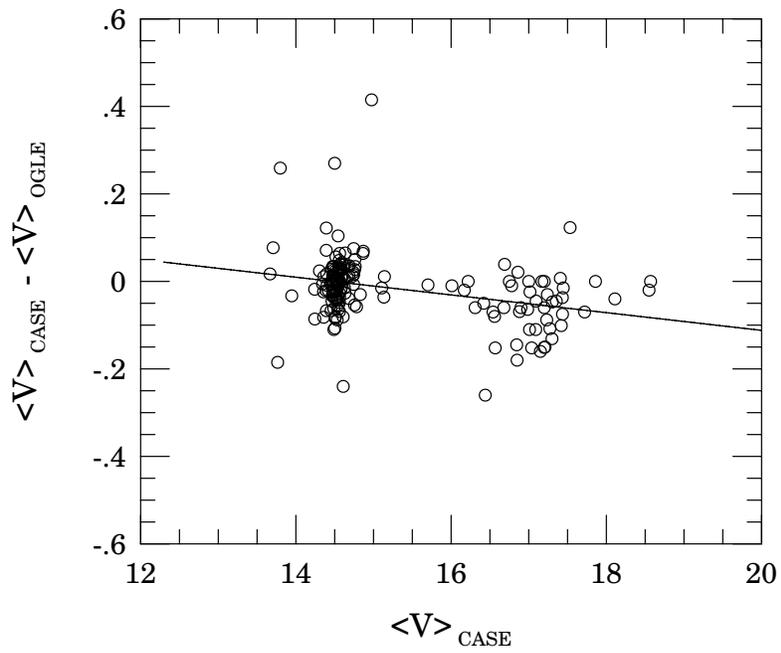}
      \caption{Comparison between our photometry and that of
Kaluzny et al. (1996, 1997a,b). Solid line corresponds to the linear
fit given in equation (3).
              }
         \label{FigVibStab}
   \end{figure*}

\clearpage

   \begin{figure*}[!h]
   \centering
\includegraphics[scale=.90]{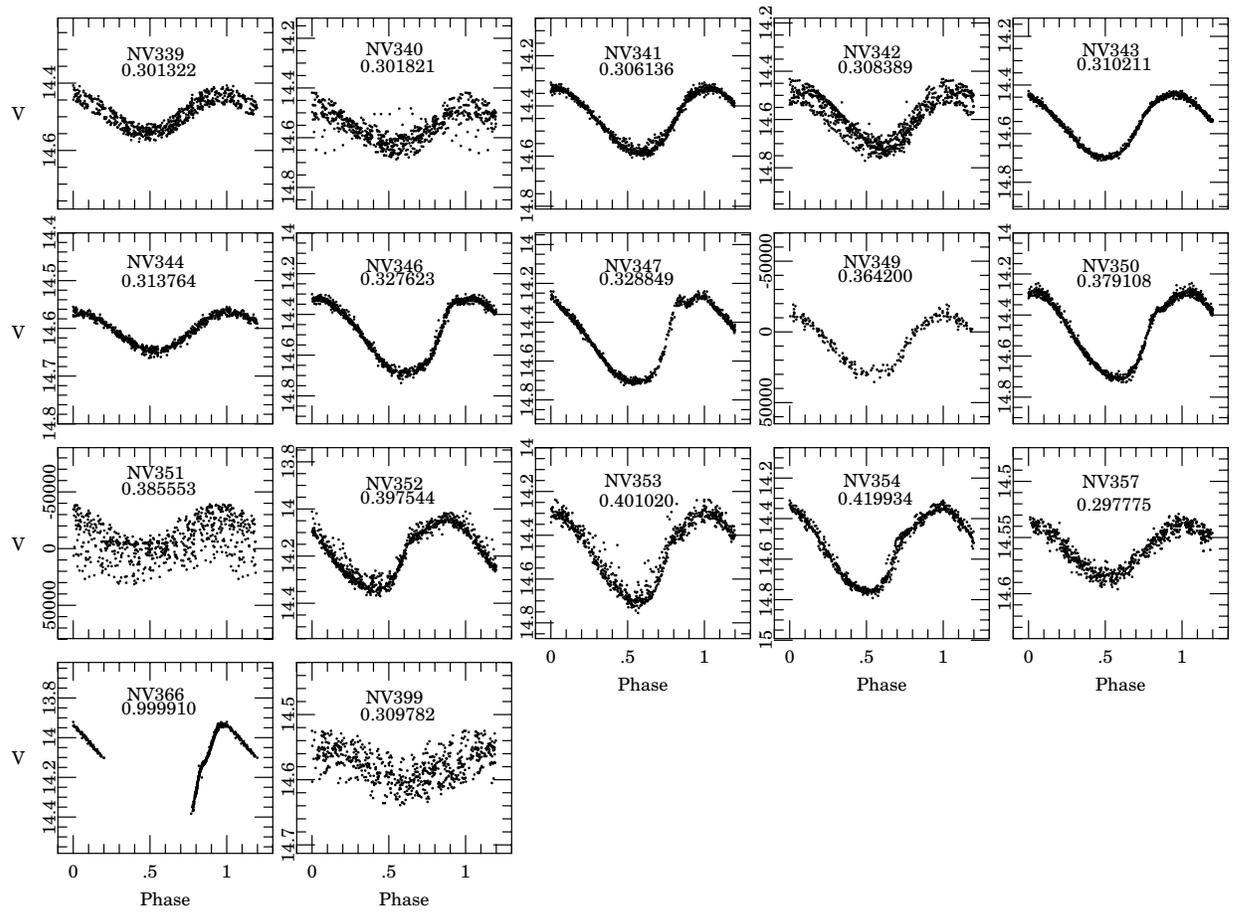}
      \caption{Light curves of newly discovered variables of RR Lyr type
              }
         \label{FigVibStab}
   \end{figure*}

\clearpage

   \begin{figure*}[!h]
   \centering
\includegraphics[scale=.90]{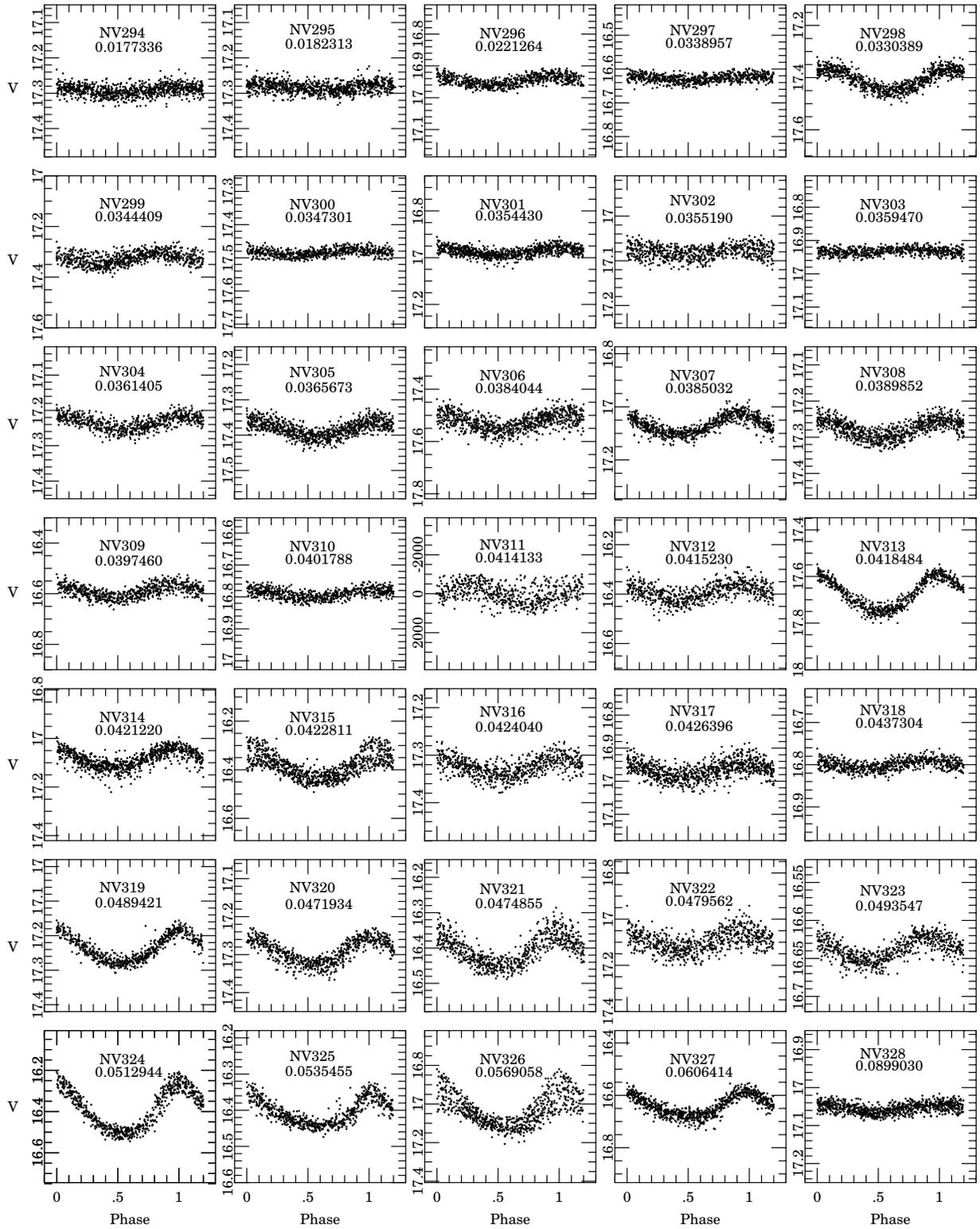}
      \caption{Light curves of newly discovered variables of SX Phe type
              }
         \label{FigVibStab}
   \end{figure*}

\clearpage

   \begin{figure*}[!h]
   \centering
\includegraphics[scale=.90]{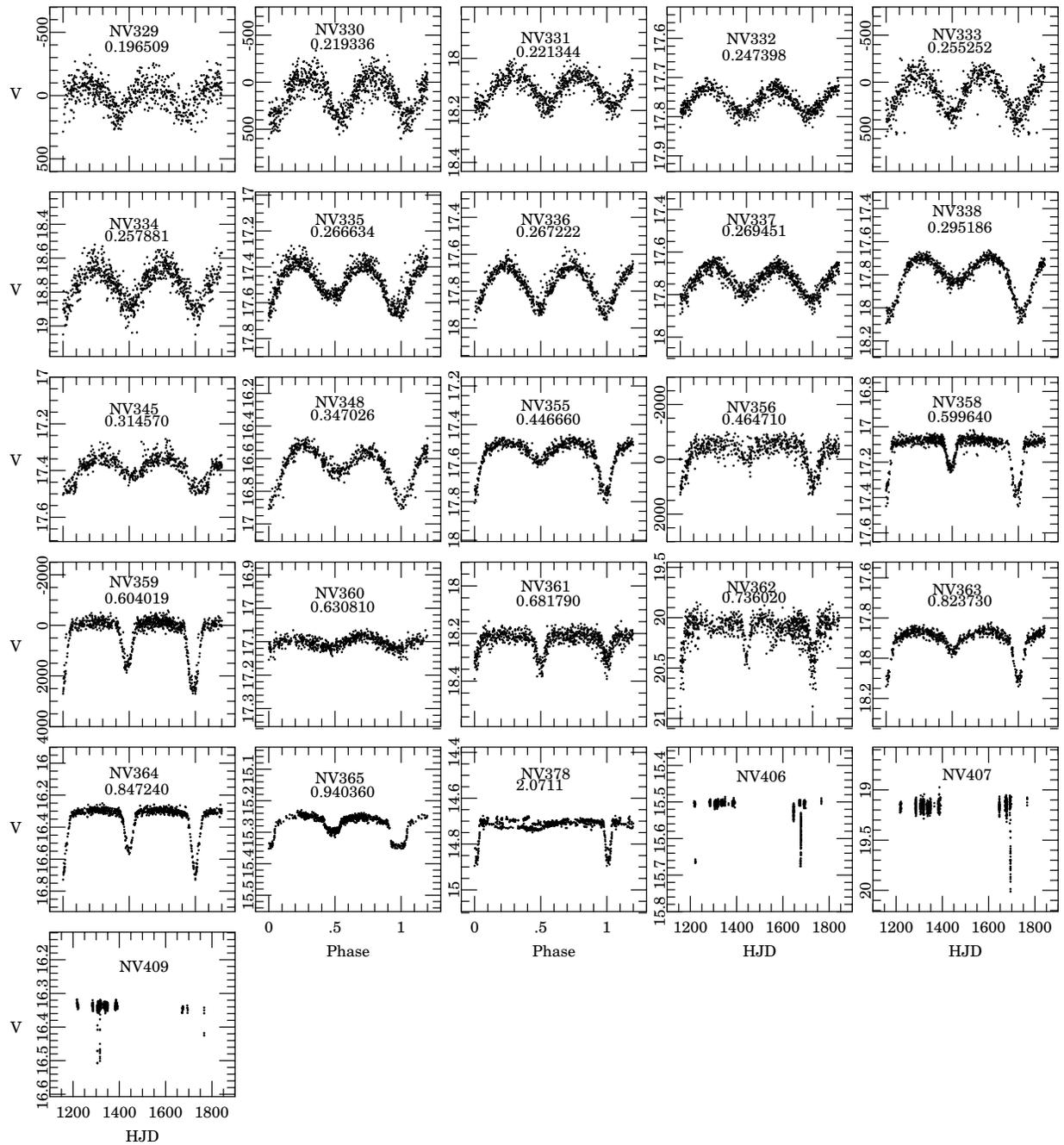}
      \caption{Light curves of newly discovered eclipsing binaries.
              }
         \label{FigVibStab}
   \end{figure*}

\clearpage

   \begin{figure*}[!h]
   \centering
\includegraphics[scale=.90]{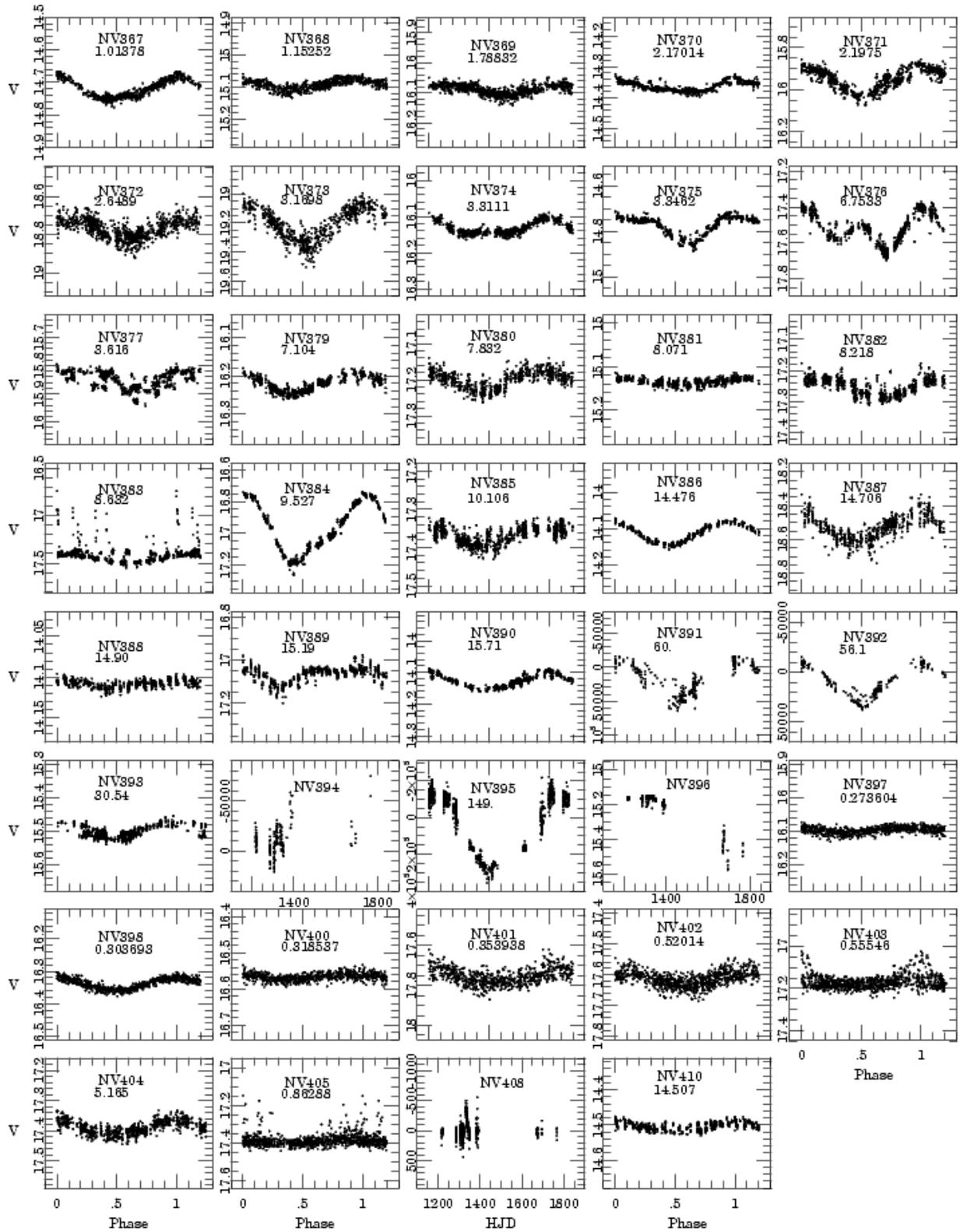}
      \caption{Light curves of newly discovered variables of types not 
               included in Figs. 4-6.
              }
         \label{FigVibStab}
   \end{figure*}

\clearpage

   \begin{figure*}[!h]
   \centering
\includegraphics[scale=.90]{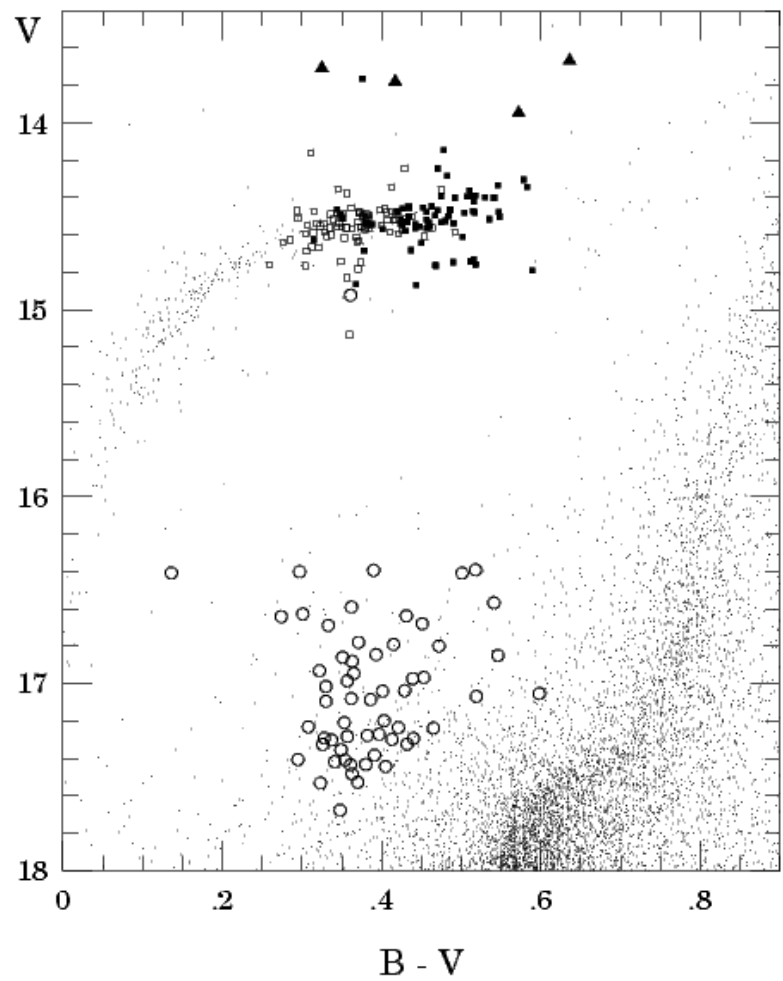}
      \caption{Color-magnitude diagram of $\omega$ Centauri with 
the positions of RR~Lyr and SX Phe stars marked. Open circles 
indicate SX Phe  stars; solid squares, RR0 stars; open squares, RR1 stars; 
triangles;  population II cepheids and BL Her stars.
              }
         \label{FigVibStab}
   \end{figure*}

\clearpage

   \begin{figure*}[!h]
   \centering
\includegraphics[scale=.90]{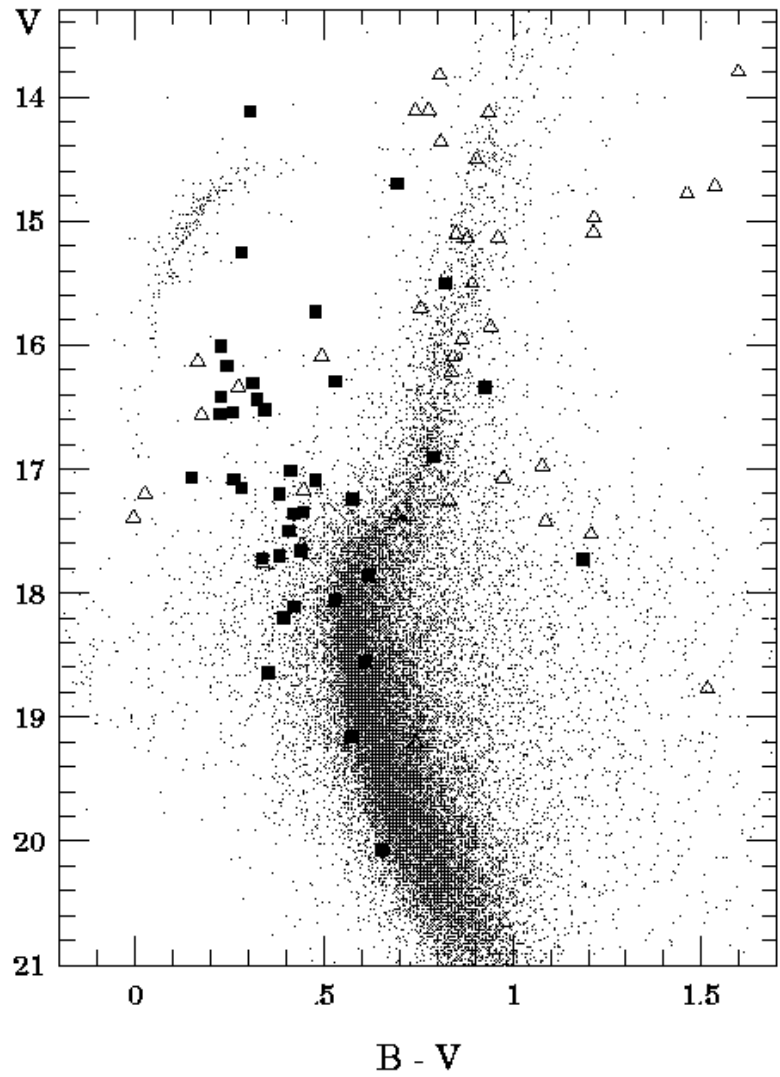}
      \caption{Color-magnitude diagram of $\omega$ Centauri with 
the positions of variables other than RR~Lyr and SX~Phe stars marked. 
Squares indicate eclipsing binaries; triangles; variables of remaining types.
              }
         \label{FigVibStab}
   \end{figure*}
\end{document}